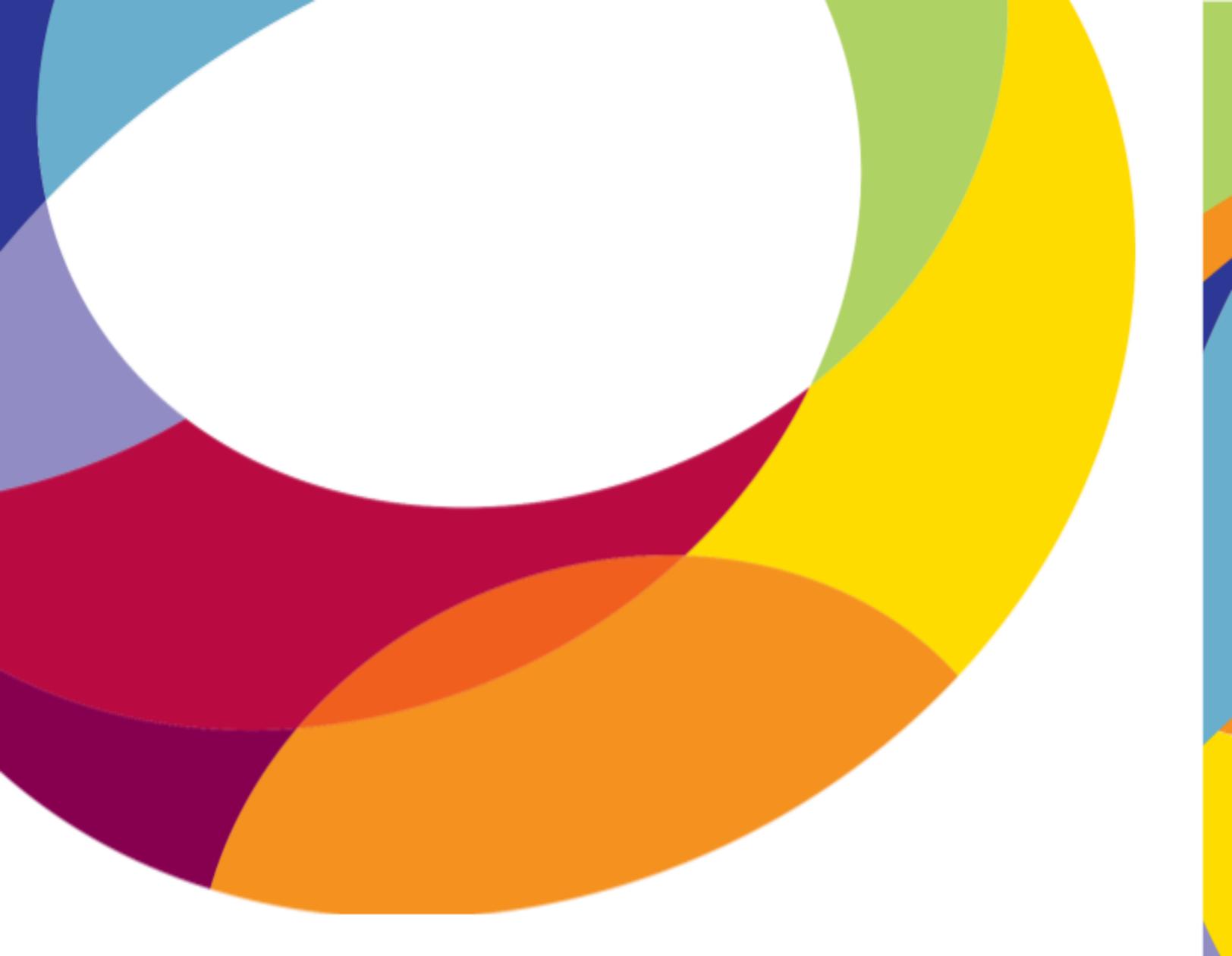

# Conference Submission and Review Policies to Foster Responsible Computing Research

**July 2024**

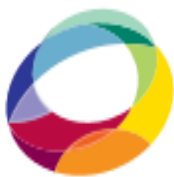

# Conference Submission and Review Policies to Foster Responsible Computing Research

CRA Working Group on Socially Responsible Computing
Subcommittee on Best Practices for Conferences

Lorrie Cranor
Carnegie Mellon University

Kim Hazelwood
Meta

Daniel Lopresti
Lehigh University

Amanda Stent
Davis Institute for AI at Colby College

**Abstract:**

This report by the CRA Working Group on Socially Responsible Computing outlines guidelines for ethical and responsible research practices in computing conferences. Key areas include avoiding harm, responsible vulnerability disclosure, ethics board review, obtaining consent, accurate reporting, managing financial conflicts of interest, and the use of generative AI. The report emphasizes the need for conference organizers to adopt clear policies to ensure responsible computing research and publication, highlighting the evolving nature of these guidelines as understanding and practices in the field advance.

**Suggested Citation:**

Cranor, L., Hazelwood, K., Lopresti, D., and Stent, A. (2024) Conference Submission and Review Policies to Foster Responsible Computing Research. Washington, D.C.: Computing Research Association (CRA).
https://cra.org/wp-content/uploads/2024/07/Report-Conference-Submission-and-Review-Policies.pdf

# EXECUTIVE SUMMARY

A growing number of computing research conferences are adding submission requirements and guidance that relate to ethical and societal considerations; however, these requirements are more common in some subfields of computing than others, and the areas they cover vary considerably. We prepared this document to aid conference organizers in developing responsible computing policies for their conferences by outlining areas where policies may be needed and highlighting issues to consider. This document reflects our understanding of responsible computing as of 2024; we expect this understanding will evolve over time and this document and conference policies will also continue to evolve. This report does not cover research integrity issues such as unethical review practices, which are discussed by the [CRA Working Group on Research Integrity](#).

Our guidelines focus on how to craft policies that promote ethical and responsible research and publication practices. We focus on the following areas:

**Avoiding harm.** We suggest that conferences include language about avoiding harm (negative consequences of research) in their submission guidelines and highlight common examples from their field that offer ways to avoid harm.

**Responsible disclosure of vulnerabilities.** Conferences that may include papers that involve disclosures of vulnerabilities in existing products or technologies may wish to offer responsible disclosure guidelines based on best practices in their field. This may include recommending or requiring submissions to either a) address the steps authors have already taken or will take to disclose vulnerabilities to stakeholders or b) explain why disclosure is not being pursued and what other steps are being taken to reduce harm.

**Ethics board review.** We suggest that conference submission guidelines state that when an Institutional Review Board (IRB) or other formal ethics review committee process is available, authors of submitted papers that qualify for that process are required to use it and explicitly document this. Submission guidelines should emphasize that regardless of whether a formal ethics review process is used, it is the responsibility of authors to follow basic principles of ethical research and to address ethical issues they become aware of even if not raised by ethics review committees.

**Obtaining consent.** We suggest that conferences require documentation of informed consent procedures or a waiver for the need of informed consent for any research that involves the collection of new data from human participants. Conferences should also require discussion of consent for research that involves reuse of previously collected data.

**Accurate reporting and reproducibility.** We suggest that conferences encourage best practices in reproducibility as appropriate to the field. This might include requesting documentation of the experimental process, requesting artifacts produced by the research to support reproducibility, or requesting reviewers to review research artifacts with an eye to reproducibility.

**Financial conflicts of interest.** Conferences may consider a range of policies to address financial conflicts of interest such as completely prohibiting submission of papers for which certain types of financial interests exist, requiring authors to disclose this as part of a submission, requiring authors to disclose it in the final paper, or leaving disclosure to the authors' discretion. Conferences should decide on an approach and offer clear guidance to authors on what is prohibited and what disclosures are required.

**Use of generative AI in computing conference publications.** While generative AI can be a useful tool for authors, it raises a number of concerns about intellectual dishonesty and ownership. Increasingly conferences and publishers are prohibiting large language models or other AI tools from being listed as co-authors but requiring them to be cited or acknowledged. In addition, some conferences and publishers have policies about the use of generative AI in the review process.

**Ways to incorporate ethics review into publication review processes.** A variety of approaches are used to incorporate ethics review into publication review processes. We recommend that conferences recruit a separate ethics review committee to review any submissions flagged for ethics review by peer reviewers. This committee can be smaller and more multidisciplinary than a regular program committee and focus its attention on the responsible research guidelines adopted by the venue.

**Reporting and retraction policies and procedures.** We suggest that conferences establish reporting and retraction policies and procedures or articulate that policies of another organization (e.g. a professional society or publisher) will be used. These policies should cover definitions of potential violations, a reporting process, an investigation process, and details about how potential outcomes (e.g., paper retraction) will be implemented.

**The growing popularity of preprint archives.** Preprint archives offer many benefits but do not provide the same level of oversight and professional responsibility as traditional publishers and conferences. We consider the use of preprint archives to be a "work in progress" when it comes to the issues of responsible computing research we have raised here.

**Tracking and transparency.** We suggest that conference organizers and/or SIGs provide high-level disclosure into the number of ethical reviews and broad categorical outcomes of reviews while keeping details confidential.

# Introduction

A 2022 National Academies Report, "Fostering Responsible Computing Research" (National Academies 2022) called on the computing research community to proactively address ethical challenges and societal concerns related to computing and computing research. Among the report's recommendations were ways to integrate ethical and societal considerations into scientific publishing. Indeed, even before the publication of the National Academies report, many computing conferences had begun adding submission requirements and guidance that relate to ethical and societal considerations; however, these requirements are more common in some subfields of computing than others, and the areas they cover vary considerably. We prepared this document *to aid conference organizers in developing responsible computing policies for their conferences* by outlining areas where policies may be needed and highlighting issues to consider. This document reflects our understanding of responsible computing as of 2024; we expect this understanding will evolve over time and this document and conference policies will also continue to evolve.

The overarching goal of responsible computing policies for conferences is to promote the publication of *socially beneficial* computing research *conducted in an ethical and responsible manner.* Our guidelines focus on how to craft policies that promote ethical and responsible research and publication practices. We do not consider a number of related issues such as plagiarism or research integrity, misconduct, what constitutes authorship, and reviewer collusion or misconduct. We also do not attempt to provide a universal definition of what "ethical" might mean in the context of computing research; some subfields have already taken such steps and we hope that the illustrative examples we highlight will inspire other subfields to develop their own standards specific to the kinds of work they do. We have not done an exhaustive search for examples or attempted to find the best examples, rather we highlight examples that have come to our attention that we believe illustrate a range of approaches that may be of interest to various subsets of the computing research community.

While the primary audience for this document is conference organizers, it may also provide guidance to computing researchers and help educate computing students about responsible research practices. However, this is not intended to be an exhaustive guide to conducting and publishing responsible computing research.

We have tried to be sensitive to the fact that practices may vary across cultures and countries, and hope that conference organizers will keep this in mind as well.

We are also sensitive to the burden that additional conference publication policies place on authors, paper reviewers, and conference organizers, all of whom are often volunteers with many other responsibilities.. Conference organizers should consider the increased reviewing burden necessary to appropriately consider all the various aspects of responsible computing research. In addition, those responsible for tenure and promotion and related processes should understand that as researchers take steps to conduct their research in a more responsible fashion, their productivity may appear to decrease when measured only in terms of publication numbers. These processes should recognize this and reward rather than penalize responsible research, and also give credit to those who participate in service activities such as reproducibility or ethics reviewing.

Our guidelines focus on the following areas: avoiding harm, responsible disclosure of vulnerabilities, ethics board review, obtaining consent, accurate reporting and reproducibility, financial conflicts of interest, use of generative AI in computing conference publications, ways to incorporate ethics review into publication review processes, reporting and retraction policies and procedures, the growing popularity of preprint archives, and tracking and transparency. Each of these areas are detailed in separate sections of this report.

## Avoiding harm

We have seen numerous instances over the recent past where computing methodologies developed out of intellectual curiosity have been employed in ways that are detrimental to individuals, specific groups, the environment, or society as a whole. Even in cases where such possible uses/abuses are impossible to predict in advance, our field is seen as holding significant responsibility (Kenneally 2012). Rather than looking for ways to absolve ourselves, we believe such energies would be better invested in fostering meaningful discussions of research ethics across all of the computing topic areas and establishing policies to put this into action. This culture is already present in some of our research communities, but it should be adopted across all of computing, just as considerations of ethics are considered inherent in areas of engineering where human safety is at risk (e.g., structural and transportation engineering), and in the study and practice of medicine. While we harbor no illusions that this will completely eliminate future situations where our work can lead to harm, instilling a sense of responsibility across all of computing and becoming a visible leader in this regard will serve our field well.

The ACM Code of Ethics (ACM 2018) defines "harm" as "negative consequences, especially when those consequences are significant and unjust." The ACM Code notes that harm is often unintentional and that "avoiding harm begins with careful consideration of potential impacts on all those affected by decisions."

The USENIX Security '25 Preliminary Call for Papers (USENIX 2024a) requires all submissions to include an ethics consideration section and discusses some ways that research can lead to harms:

> Without sufficient precautions, research endeavors can lead to negative outcomes. People or other entities, like companies, might experience negative outcomes during the research process itself, immediately after the research is published, or in the future. These negative outcomes might be in the form of tangible harms (e.g., financial loss or exposure to psychologically disturbing content). Or, these negative outcomes could be violations of human rights even if there are no directly tangible harms (e.g., the violation of a participants' right to informed consent or the violation of users' right to privacy via the study of data that users expect and desire to be private). Further, due to the complexity of today's computing systems, people could experience these negative outcomes either directly or indirectly in unexpected ways (see The Menlo Report).

We expect authors to carefully and proactively consider and address potential negative outcomes associated with carrying out their research, as well as potential negative outcomes that could stem from publishing their work. Failure to do so may result in rejection of a submission regardless of its quality and scientific value.

Although causing negative outcomes is sometimes a necessary and legitimate aspect of scientific research in computer security and privacy, authors are expected to document how they have addressed and mitigated the risks. This includes, but is not limited to, considering the impact of the research on deployed systems, understanding the costs the research imposes on others, safely and appropriately collecting data, considering the well-being of the research team, and following ethical disclosure practices.

In addition, the USENIX Security call for papers refers to a separate USENIX Security '25 Ethics Guidelines document (USENIX 2024b) that offers further background and guidance on ethical issues with examples relevant to the security field.

Conferences might also consider offering guidance on avoiding societal or environmental harms, for example, harms related to increased energy consumption (Richard 2021), threats to sustainability, and negative impacts on the workforce.

Conferences should include language about avoiding harm in their submission guidelines and, as appropriate, highlight common examples from their field that offer ways to avoid harm.

## Responsible disclosure of vulnerabilities

For conferences where papers may include vulnerability disclosures (e.g., software vulnerabilities in a particular program, design weaknesses in a hardware system, or any other kind of vulnerability in deployed systems), submission guidelines should emphasize the need to act in a way that avoids harm and, where possible, affirmatively protect those impacted. This will often involve disclosing the vulnerability to vendors of affected systems (or organizations that coordinate vulnerability disclosures) and other stakeholders with sufficient time for remediation to occur prior to publication. Conferences may wish to offer responsible disclosure guidelines based on best practices in their field. Submissions should address the steps authors have already taken to disclose vulnerabilities and their plan and timeline for disclosure if disclosure has not already occurred. In the event that disclosure is not appropriate or may cause more harm, submissions should address why disclosure is not being pursued and other steps that are being taken to reduce harm. In all cases, submissions should be encouraged to discuss the risks and benefits associated with the research and its publication.

**Examples:**
- The IEEE Symposium on Security and Privacy 2025 includes a disclosure expectation with a recommended timeline (IEEE S&P 2024a):
  > Where research identifies a vulnerability (e.g., software vulnerabilities in a given program, design weaknesses in a hardware system, or any other kind of vulnerability in deployed systems), we expect that researchers act in a way that avoids gratuitous harm to affected users and, where possible, affirmatively protects those users. In nearly every case, disclosing the vulnerability to vendors of affected systems, and other stakeholders, will help protect users. It is the committee's sense that a disclosure window of 45 days https://vuls.cert.org/confluence/display/Wiki/Vulnerability+Disclosure+Policy to 90 days https://googleprojectzero.blogspot.com/p/vulnerability-disclosure-faq.html ahead of publication is consistent with authors' ethical obligations.
  >
  > Longer disclosure windows (which may keep vulnerabilities from the public for extended periods of time) should only be considered in exceptional situations, e.g., if the affected parties have provided convincing evidence the vulnerabilities were previously unknown and the full rollout of mitigations requires additional time. The authors are encouraged to consult with the PC chairs in case of questions or concerns.
  >
  > The version of the paper submitted for review must discuss in detail the steps the authors have taken or plan to take to address these vulnerabilities; but, consistent with the timelines above, the authors do not have to disclose vulnerabilities ahead of submission. If a paper raises significant ethical and/or legal concerns, it will be checked by the REC and it might be rejected based on these concerns. The PC chairs will be happy to consult with authors about how this policy applies to their submissions.

- Similarly, the USENIX Security '25 Ethics Guide includes the following, but without a specific timeline (USENIX 2024b):
  > …absent strong and convincing reasons otherwise, we expect researchers to disclose vulnerabilities [to vendors or organizations that coordinate vulnerability disclosure] as soon as they are discovered. If the researchers believe that a different timeline is the most ethical in their situation, they should present clear and convincing arguments for that different timeline. The arguments should clearly articulate why a delayed disclosure is in the best interest of users or people in general, e.g., most supportive of these people's wellbeing or least likely to violate their rights. Submissions that fail to disclose prior to submission and that do not present convincing ethical arguments for delaying disclosure may be rejected or may receive a revision decision.

We note that, in certain cases the right decision to avoid harm may be *not* to publish certain results until efforts have been made to successfully mitigate the issues that are the focus of the work.

## Ethics Board Review

Conference submission guidelines should ask that authors of submitted papers weigh the risks and benefits of the research and document steps taken to minimize risk. In addition, where an Institutional Review Board (IRB) or other formal ethics review committee process is available, authors of submitted papers that qualify for that process should be required to use it and explicitly document this. Where no such formal process is available, authors might be encouraged to systematically consider ethical concerns (possibly following similar processes used in another country or organization where formal processes exist), ideally inviting someone not involved in their project to ask questions and offer feedback to help surface issues the research team may not have considered.  A conference or professional organization might recruit a pool of experts who can consult with authors on these topics, as some organizations have done (e.g., Northeastern 2023, University of Pittsburgh 2023, University of Utah 2023). Submission guidelines should emphasize that regardless of whether an IRB or other ethics review process is used, it is the responsibility of authors to follow basic principles of ethical research and to realize that as subject matter experts, they may become aware of ethics issues not raised by ethics review committees.

**Examples:**
- The NeurIPS 2023 code of ethics (NeurIPS 2023) states:
    > …if the research presented involves direct interactions between the researchers and human participants or between a technical system and human participants, authors are required to follow existing protocols in their institutions (e.g., human subject research accreditation, IRB) and go through the relevant process. In cases when no formal process exists, they can undergo an equivalent informal process (e.g., via their peers or an internal ethics review).
- The USENIX Security '25 Ethics Guidelines (USENIX 2024) states:
    > …if the submitted research has potential to create negative outcomes and authors have access to an Institutional Review Board (IRB), then authors are encouraged to consult this IRB and document its response and recommendations in the paper. In some parts of the world, and in some situations, consulting with the IRB may be required. IRBs are not, however, expected to understand computer security research well or to know about best practices and community norms in our field, and so IRB approval does not absolve researchers from considering ethical aspects of their work. In particular, IRB approval is not sufficient to guarantee that the PC will not have additional concerns with respect to potential negative outcomes associated with the research. Hence, the discussion of IRB approval (if relevant) will likely only be a subset of the ethics discussion. (If authors do not have access to an IRB but are doing human subjects-related research for which IRB

approval might be required elsewhere and of other researchers, then the authors are encouraged to explicitly state that they do not have access to an IRB and, instead, focus on what mechanisms they used as they considered and addressed ethical considerations.)

## Obtaining Consent

Informed consent is permission given by a person providing (or permitting collection or use of) data or permitting an intervention, made with full knowledge of the possible consequences of their action.

The reason we seek informed consent for research and development activities involving humans is to ensure that the rights and welfare of individuals are protected. Consideration of individual rights includes consideration of the right to life, liberty, privacy, and one's own property (including intellectual property). Consideration of individual welfare includes the prevention of discrimination and harassment and the support of individual health, security, and happiness. Research ethics rules in many countries include the principle that, in general, no amount of perceived benefit to society is worth denying the rights or welfare of individuals (as in the case of Henrietta Lacks(Lang 2023)) or groups (as in the case of the Tuskegee syphilis experiments (Waxman 2017)).

Computing research that involves data from or pertaining to individuals, or that is applied as interventions affecting individuals, should not be undertaken without obtaining informed consent. We acknowledge that this may be difficult in some cases: for example, when collecting data from the web it may be hard or impossible to contact those whose data is being collected; or it may be hard or impossible to help non-technologists understand reasonably foreseeable possible consequences of use of a computing system (Fiesler 2022).

In some contexts, the risks of not seeking informed consent are minimal, or are outweighed by the benefits. Governments, universities, and companies around the world have established procedures for the review of proposed research that are intended to address questions of whether, when, and how to obtain informed consent for research. In addition, there is a growing body of regulation around informed consent in responsible applications of computing. For example, several states now have regulations governing the creation and application of facial recognition technology or personalization / recommender systems. However, research review boards are often staffed by people who may not have expertise in the risks and benefits of large data or of distributed computing systems with large user bases. Furthermore, research review boards are generally not set up to address other questions around fair and responsible use of data from individuals, such as protection of intellectual property (Abbott 2022, Levendowski 2018). In addition, regulation often lags behind technological development.

Conference organizers should require documentation of research review for any research that involves the collection of new data from human participants or the application of new technology to human participants. This documentation should include review of informed consent procedures or a waiver for the need of informed consent. Conference organizers should also require discussion of consent for research that involves reuse of previously collected data, as individuals may consent for their data to be used in a particular context, but not for it to be redistributed or used in other contexts.  If a community identifies a certain dataset as being problematic – for example, having been assembled at a time before the concerns we raise here were taken into consideration – conference organizers and reviewers can flag it and offer alternative suggestions. Research communities may find it necessary to devote resources toward building new datasets that satisfy the goals of socially responsible computing.

As one example, the ACL Rolling Review Responsible Research Checklist (ACL 2022b) asks authors to identify the source of any data, and to address whether use of the data is justified because the data is in the public domain, licensed for research purposes, used with the consent of its creators or copyright holders, or otherwise justified. It also asks authors whether their use of existing datasets was consistent with the intended use of those datasets. Finally, it asks authors to discuss whether and how consent was obtained from human participants, and whether data collection protocols were approved, or determined exempt, by an ethics review board.

## Accurate Reporting and Reproducibility

A core tenet of the scientific method is integrity in the process and documentation of scientific research, including documenting and tracking experiments and artifacts (such as code, data and models). Documentation is important to ensure reproducibility and replicability, which are important to ensure accuracy and objectivity. We note that there are many inconsistencies in the use of the terms *reproducibility* and *replicability.* A National Academies report (NAS 2019) has adopted these definitions: "reproducibility is obtaining consistent results using the same input data; computational steps, methods, and code; and conditions of analysis…. Replicability is obtaining consistent results across studies aimed at answering the same scientific question, each of which has obtained its own data." Failure to ensure accurate reporting and reproducibility can have negative impacts on not just the research field itself but also society as a whole. For example, decades of research on Alzheimer's may have been based on a faulty paper (Piller 2022).

In the field of computing research, Raff attempted to reproduce results from 255 machine learning papers published over four decades and discovered that only two thirds were fully reproducible. Raff also found that whether or not the papers' authors released their code has no significant relationship with the paper's independent reproducibility (Raff 2019).  In the same year, Bouthillier et al. highlighted that inherent randomness in machine learning experiments could have a greater impact on reported results than true differences in the methods under study (Bouthillier et al. 2019).

Conference organizers should encourage best practices in reproducibility (e.g., Stodden et al. 2016) as appropriate to the field. This might include:
1. **Requesting documentation of the experimental process.** For example, organizers may request preregistration (Center for Open Science 2023), or documentation about how many runs were done for a simulation, or how many hyperparameter settings were explored for model training (Dodge et al. 2019, Hill 2019, David et al. 2019).
2. **Requesting authors to provide artifacts produced by the research to support reproducibility.** For example, organizers may request that authors provide (documented and licensed) code (Arvan 2022, Pineau 2021), models (Mitchell 2019), and/or data (Gebru 2021, Feinberg 2020). In some cases it might be unwise or impossible to release research artifacts (for example, when doing so would violate informed consent); in these cases, organizers may allow authors to instead share artifacts in a controlled setting for the purposes of review only (e.g., Fröbe 2023) or ask authors to document reasons for not providing research artifacts.
3. **Requesting reviewers to review research artifacts with an eye to reproducibility.** The mere provision of research artifacts does not amount to reproducibility (for example, if code is not documented and runnable). Review of research artifacts may be done as part of the regular submission review process or only for submissions deemed above a bar following initial review. Review of research artifacts might be done by a separate pool of reviewers (e.g., PETS 2023, USENIX 2025a), with relevant expertise, such as subject matter expertise relevant to the conference, software engineering, and statistics. As review of research artifacts creates extra workload on review committees, committees may need to be expanded in order to take this on.

To facilitate reproducible computing research, conference organizers may encourage authors to provide documentation and artifacts as appendices or supplemental materials to a paper submission. Wherever possible, supplemental materials should be retained *by the conference*, as several studies have shown that author-hosted artifacts disappear from the internet very quickly (Arvan 2022).

While reproducibility is an important goal with respect to experimental research (including computational simulations) and engineering research (papers describing the implementations of systems), there are types of computational research where reproducibility is less important. If conference organizers ask authors to address reproducibility, authors should be able to decline if they can provide appropriate justification. Furthermore, reproducibility may be limited when hardware or software becomes obsolete; generally, authors should not be expected to provide updates, support, or otherwise make provisions for research artifacts to continue to function far into the future.

**Examples:**
- The NeurIPS code of ethics (NeurIPS 2023) says:
  > Any work submitted to NeurIPS should be accompanied by the information sufficient for the reproduction of results described. This can include the code, data, model weights, and/or a description of the computational resources needed to train the proposed model or validate the results.
- The USENIX Security '25 Preliminary Call for Papers (USENIX 2024a) says:
  > Authors are expected to openly share their research artifacts by default. This initiative is part of a broader commitment to foster open science principles, emphasizing the sharing of artifacts such as datasets, scripts, binaries, and source code associated with research papers. If, for some reason (such as licensing restrictions), artifacts cannot be shared, a detailed justification must be provided. Artifacts need to be available for the Artifact Evaluation committee after paper acceptance and before the final papers are due.
- Dodge et al. (2019) include a short checklist for reporting experimental computational research with reproducibility in mind.
- SIGPLAN also has a checklist for empirical evaluation to address replicability and reproducibility (Berger et al. 2019).

# Financial Conflicts of Interest

Ethical concerns may be raised when authors have a real or perceived financial interest in their results and in having them published. This may occur, for example, when an author submits a paper that evaluates a product or component being sold or expected to be sold by a company they work for or have a stake in. This may also occur when an author has received research funding from a company and used it to evaluate that company's products. In some cases, this financial conflict of interest may call into question whether the evaluation in the paper is actually fair and unbiased. In other cases the authors may present evidence of steps they took to minimize potential bias or may argue that the component being evaluated has been released as open source and is used by competitors as well and thus the potential conflict may be mitigated.

There may be competing values to be considered in such situations. We would like to avoid having the peer review process endorse papers that may have biased results and primarily contribute to a company's profit rather than scientific knowledge. However, we would also like to increase transparency about how technology works and how it has been evaluated. Thus, we may want to encourage papers about products to be submitted for peer review.

Conferences may consider a range of policies to address financial conflicts of interest. For example, they may completely prohibit submission of papers for which certain types of financial interests exist. Alternatively, they may require authors to disclose this as part of a submission or in the final paper, or they may leave disclosure to the authors' discretion. In the event that disclosures are required on submission, they might be evaluated separately from the normal review process or by the paper reviewers to determine whether a conflict of interest should disqualify the paper from publication. If handled by the reviewers, it may be impossible to hide the identity of the author or their organization from the reviewer, but some conferences may decide that is a tradeoff they are willing to make.  Conferences should decide on an approach and offer clear guidance to authors on what is prohibited and what disclosures are required.

Examples:
- The ACM FAccT 2023 conference required  authors to provide a detailed funding disclosure in their camera-ready papers but asked that these statements not be included in submitted papers in order to preserve mutually anonymous review (ACM FAccT 2023).
- The PLOS ONE journal requires authors to make a "Financial Disclosure Statement" as part of their submission, but not part of their manuscript. The statement must include a variety of details such as grant numbers, companies that funded the study or authors, a statement about whether any sponsors or funders other than the named authors played any role in the research or manuscript preparation, details about related patents or patent applications, and details about products released or under development based on the research.  PLOS ONE provides a detailed definition of "Competing Interests" that must also be disclosed as part of submissions. PLOS ONE includes a "Funding Statement"  in the metadata of each published article (PLOS Medicine Editors 2008, PLOS ONE 2024).
- The 46th IEEE Symposium on Security and Privacy has a financial conflicts policy which defines both financial and non-financial competing interests and states (IEEE S&P 2024b):
    > Authors need to include a disclosure of relevant financial interests in the camera-ready versions of their papers. This includes not just the standard funding lines, but should also include disclosures of any financial interest related to the research described. For example, "Author X is on the Technical Advisory Board of the ByteCoin Foundation," or "Professor Y is the CTO of DoubleDefense, which specializes in malware analysis."

## Use of Generative AI in CS Conference Publications

There is much current discussion regarding the use of generative AI in scientific publications. Springer-Nature has recently declared, for example, that large language model tools (LLMs) cannot be listed as co-authors (Nature 2023) and that the use of LLMs must be acknowledged in the paper. AAAI, ACM, and IEEE have taken similar stands (AAAI 2023, ACM 2023b, IEEE 2024b). IEEE's guidelines also state that all authors must approve of the article as accepted for publication or agree to be held accountable for any issues relating to the correctness or integrity of the work, terminology which seems to rule out designating an LLM as a co-author (IEEE 2023).

While we expect that conference organizers will take some of their cues from publishers as above, authorship is not the only way in which generative models might be employed. As this is a rapidly evolving area, broad consensus may take some time to develop.  Here we list some of the ethical issues that concern us and for which best practices seem to be necessary. First would be the use of LLMs to "improve" existing writing, in a way that can be seen as building on existing tools like spelling and grammar checkers. While such uses may appear innocuous, the degree to which LLMs can "appropriate" a particular author's style may create the potential for copyright violations, or at least some new form of intellectual dishonesty (we might call this "inadvertent plagiarism"). A more insidious turn of events could arise if generative AI were used to propose experiments to support a current line of research, or to generate "new" hypotheses to consider and act on. In cases like these, the researchers using the AI might be oblivious to the actual origins of the ideas (which might come from the training data) and fail to acknowledge the intellectual contributions of colleagues used in training the AI. This could lead to a significant increase in post-publication challenges that would have to be adjudicated after the fact, since the current peer-review paradigm seems ill-equipped to handle the added burden of this sort of verification.

This new responsibility falling on authors will require extra diligence regarding the source of the ideas they include in their research. It seems unlikely, at least for now, that generative AI can be trusted to provide an accurate accounting for the source(s) of an idea it synthesizes, so authors must track down these sources. In addition, while conference organizations and publishers have existing policies on plagiarism, AI-induced plagiarism may require new policies and procedures. Conferences should consider at minimum explicitly requiring that authors cite the use of any AI tools.

Some conferences and publishers are also adding policies about the use of generative AI as part of the review process to address concerns about relying on AI to make decisions as well as concerns about leaking information about submissions, authors, or the confidential review process. For example, the ACM peer review policy prohibits reviewers from uploading submissions into "any system managed by a third party, including LLMs, that does not promise to maintain the confidentiality of that information by reviewers, since the storage, indexing, learning, and utilization of such submissions may violate the author's right to confidentiality." However, ACM allows reviewers to "use generative AI or other third-party tools with the sole purpose of improving the quality and readability of reviewer reports for the author, provided any and all parts of the review that would potentially identify the submission, author identities, reviewer identity, or other confidential content is removed prior to uploading into third party tools" (ACM 2023a). IEEE goes a step further in their journal and magazine reviewer guidelines and prohibits reviewers from using AI tools to help write reviews of IEEE articles in any way (IEEE 2024a).

# Ways to Incorporate Ethics Review into Publication Review Processes

The National Academies Report (NAS 2022) calls for review committees to incorporate review of the ethical and societal implications of research into the publication review process. We call this "ethics review" herein, but we mean to include reviewing with an eye to all aspects of responsible computing including the topics discussed in earlier sections. Today, conferences do this in one of three ways:

1. Ask reviewers to consider ethics review during the regular peer review process. This has the advantage of simplicity, but also significant disadvantages. For example, reviewers may not be qualified to assess various ethics review criteria or may not want to take time to review research artifacts accompanying a submission (such as code, data, or models).
2. Ask area chairs, senior area chairs, or decision editors to conduct ethics reviews. This has the advantage that these experts are likely to be more senior than regular reviewers and program committee members, but also may impose a large burden on them if they handle a large number of papers that all require ethics reviews.
3. Recruit a separate ethics review committee. The ethics review committee reviews any submissions flagged for ethics review by peer reviewers. This committee can be smaller and more multidisciplinary than a regular program committee and focus its attention on the responsible research guidelines adopted by the venue.

In all cases, reviewers need to be educated about the responsible research guidelines adopted by the venue; however, in Approach 3, only the ethics review committee requires in-depth education. We recommend this approach, which has been explored by NeurIPS (Luccioni et al. 2022) and ACL Rolling Review (ACL 2022a). Although some submissions that should undergo enhanced ethics review will not be flagged by the regular pool of peer reviewers, those that are flagged can be reviewed by a diverse committee of experts in human subjects research, privacy/security, philosophy, and so on. This committee may be a standing committee, well-versed in the responsible research guidelines adopted by the venue. The ethics review process can take place after the regular review process (papers flagged for ethics review but otherwise marked for acceptance may be conditionally accepted).

A concern that arises when extra steps like this are introduced into the review process is the potential for abuse. If some papers are flagged for a more detailed review, this could introduce an extra delay in providing feedback to authors and ultimately rendering a decision on the paper. This could disadvantage some authors over others, so careful attention must be paid to timelines, including the motivation behind those raising objections to a work.

## Reporting and Retraction Policies and Procedures

Reporting and retraction policies and procedures should be in place and easily accessible for every conference, or should leverage and clearly articulate the broader policies and procedures that apply, such as the ACM, IEEE, or SIG guidelines.

Policies and procedures should cover: general definitions to clarify potential violation scenarios, a process for reporting potential violations (e.g., ACM 2020), an investigation process (e.g., ACM 2023c), and processes for each of the potential outcomes. For instance, in the case where the outcome is paper retraction, a clear process on *how* to retract papers once that decision has been made should be provided (e.g., ACM 2022) and customized for the individual venue. Furthermore, in cases where the policy dictates that violations result in notification to an individual's employer, this process should also be articulated.

If an investigation is required, it is important that it be conducted in a thorough but timely fashion; unnecessary delays in rendering a decision – especially when authors are found to be innocent of violating ethical guidelines – is potentially an ethical violation in itself.

## The Growing Popularity of Preprint Archives

An additional complication arises as a result of the growing popularity of preprint archives. Such archives are aimed at addressing several laudable goals, including accelerating the speed with which new results can be disseminated, broadening accessibility, and lowering the costs of publication. At the same time, current models, which adopt a crowd-sourcing paradigm, do not provide the same level of oversight and professional responsibility as traditional publishers, journal editorial boards, and conference program committees. We consider the use of preprint archives to be a "work in progress" when it comes to the issues of responsible computing research we have raised here.

## Tracking and Transparency

While the details and specifics of any ethics reviews are confidential, conference organizers and/or SIGs should provide high-level visibility into the number of ethical reviews and broad categorical outcomes of reviews (e.g., retractions). Disclosure of this high-level data should be a mandatory requirement for sponsorship by professional organizations.

# Conclusions

These guidelines represent a snapshot in time. As we collectively gain more experience, updates will undoubtedly be necessary. Efforts to develop and share best practices for conference submission and reviewing are a key step toward our field assuming greater responsibility for our research and its impact on society.

*We expect this document will need to be revised over time. Please send information about additional steps conference committees are taking to promote responsible computing and suggestions for revisions to: conference-policies-report@cra.org.*

# Acknowledgments

We thank Alex Wolf, Stephanie Forrest, and members of the CRA Board for providing helpful feedback on a draft version of this report.